\documentclass[12pt]{article}
\usepackage{graphicx,amssymb,psfrag,fullpage}

\usepackage{amsmath}

\newcommand{\ie}{{\it i.e.}}
\newcommand{\Expect}{\mathop{\bf E{}}}

\newcommand{\QED}{~~\rule[-1pt]{6pt}{6pt}}
\newcommand{\reals}{{\mbox{\bf R}}}

\newcommand{\symm}{{\mbox{\bf S}}}  

\newcommand{\Tr}{\mathop{\bf Tr}}

\newcommand{\BEAS}{\begin{eqnarray*}}
\newcommand{\EEAS}{\end{eqnarray*}}
\newcommand{\BEA}{\begin{eqnarray}}
\newcommand{\EEA}{\end{eqnarray}}
\newcommand{\BEQ}{\begin{equation}}
\newcommand{\EEQ}{\end{equation}}
\newcommand{\BIT}{\begin{itemize}}
\newcommand{\EIT}{\end{itemize}}
\newcommand{\BNUM}{\begin{enumerate}}
\newcommand{\ENUM}{\end{enumerate}}

\newcommand{\BA}{\begin{array}}
\newcommand{\EA}{\end{array}}

\begin{document}

\title{Static versus dynamic arbitrage bounds on multivariate option prices}
\author{Alexandre d'Aspremont \thanks{EECS Dept., U.C. Berkeley, CA 94720. alexandre.daspremont@m4x.org}} \maketitle

\begin{abstract}
We compare static arbitrage price bounds on basket calls,
\ie~bounds that only involve buy-and-hold trading strategies, with
the price range obtained within a multivariate generalization of
the \cite{Blac73} model. While there is no gap between these two
sets of prices in the univariate case, we observe here that
contrary to our intuition about model risk for at-the-money calls,
there is a somewhat large gap between model prices and static
arbitrage prices, hence a similarly large set of prices on which a
multivariate \cite{Blac73} model cannot be calibrated but where no
conclusion can be drawn on the presence or not of a static
arbitrage opportunity.
\end{abstract}

\section{Introduction}
In the classic unidimensional \cite{Blac73} framework, there is no
gap between the range of model prices on one hand and the range of
prices that create a buy-and-hold arbitrage opportunity on the
other. This means in practice that if we can't extract a market
implied volatility from the price of a call, then we know that
there either an error in the data or a buy-and-hold arbitrage
opportunity in the market that makes this price unviable on the
long run.

The \cite{Blac73} call price formula is a strictly increasing
function of the volatility and can be easily inverted, we also know
from \cite{Laur00} for example that testing for the presence of a
buy-and-hold arbitrage is equivalent to testing for positivity,
monotonicity and convexity on call prices, which can be done by
inspection. The unidimensional setting is then extremely favorable,
since model calibration can be achieved at very little numerical
cost, and a final answer is obtained as easily on the presence or not
of buy-and-hold (or static) arbitrage opportunities. Because of these
numerical properties and for consistency, call prices are always
quoted in terms of their implied volatility.

Unfortunately, those key numerical properties are lost in a
multivariate setting. As we will see below, calibrating a
multivariate \cite{Blac73} model is a non trivial exercise and
testing for the absence of static arbitrage between basket options
becomes an NP-Hard problem (see \cite{bert00}).

Equity derivatives markets don't make very frequent use of basket
options beyond some elementary spread options. However, interest
rate derivatives market volatility information is mostly
concentrated in caps and swaptions, which can be seen as basket
options on forward rates (see \cite{Rebo98} or \cite{dasp02d}
among others). While empirical evidence suggests that multivariate
lognormal approximations to market models (see \cite{Brac99} or
\cite{dasp02d}) calibrate very well to prices of caps and
swaptions, a satisfactory joint model of correlation and smile
features has yet to be designed. This means in practice that there
remains a range of option prices on which these models cannot be
calibrated but where no conclusion can be drawn on the presence or
not of a static arbitrage. In this work, we try to quantify the
magnitude of this range of prices.

The paper is organized as follows, in section two we show how to
derive bounds on the price of a basket call within a multivariate
lognormal model. In section three we detail various relaxation
techniques to compute static arbitrage bounds on the price of
baskets while section four details some numerical results.

\section{Model price bounds}
In this section, we compute the range of prices covered by a
multivariate lognormal model when calibrated to a set of (liquid)
market instruments. This is a hard numerical problem in general but
excellent estimates of these bounds can be computed in a multivariate
\cite{Blac73} model. We briefly describe the results of
\cite{dasp02d} in a simple equity setting, for more details on market
model approximations and interest rate options pricing, we refer the
reader to \cite{Rebo98}, \cite{Brac99} or \cite{dasp02d}.

In this setting, the dynamics of the assets $F_t^i$ are given by:
\[
dF_{s}^{i}=F_{s}^{i}\sigma^{i}dW_{s},
\]
where $W_{s}$ is a $n$-dimensional Brownian motion and
$\sigma^{i}\in \reals^{n}$ for $i=1,\ldots ,n$ are the volatility
parameters. We shall denote by $X \in \symm^n$ the corresponding
covariance matrix, with $X_{ij}=\sigma^{iT}\sigma^{j}$. The sum of
lognormally distributed assets is not lognormal, but we sue a
lognormal approximation to price basket calls. From
\cite{dasp02d}, we know that the price of a basket call with
payoff:
\[
\left( \sum_{i=1}^n{w_iF_T^i-K}\right)^+
\]
at time $T$, can be approximated by a \cite{Blac73} call price using
an appropriate variance $V_{T}$ such that:
\BEQ
C=BS(w^TF_{t},K,T,V_{T})=(w^TF_{t})\mathcal{N}(h(V_{T}))-K
\mathcal{N}\left( h(V_{T})-\sqrt{V_{T}}\right), \label{e-bs-price}
\EEQ
where $\mathcal{N}(x)$ is the CDF of the normal distribution:
\[
h\left( V_{T}\right) =\frac{\left( \ln \left( \frac{w^TF_{t}}{K}
\right) + \frac{1}{2}V_{T}\right) }{\sqrt{V_{T}}},
\]
with
\[
V_{T}=\Tr\left( \Omega X\right)T,
\]
where $\Omega\in\symm^n$ is the matrix given by
$\Omega=\hat{w}\hat{w}^{T}$ and
\[
\hat w_{i}=\frac{w_iF_t^i}{w^TF_{t}}.
\]
This gives swaptions prices accurate to within $1$-$4$ basis
points in the Libor market model. Furthermore, a good estimate of
the hedging tracking error can be computed by robustness argument
(see \cite{dasp02b}).

Since the \cite{Blac73} formula is strictly increasing with its
variance term $V_{T}$, computing bounds on the model price of a
basket call given market price data on other baskets (with the same
maturity) is equivalent to solving the following semidefinite
program:
\BEQ
\label{e-sdp-bs}
\BA{ll}
\mbox{max./min.}  & \Tr(\Omega_{0}X) \\
\mbox{subject to} & \Tr(\Omega_{i}X)=V_{T,i},\quad
i=1,\ldots,m\\
                  & X \succeq 0,
\EA
\EEQ
in the variable $X\in\symm^n$ with $V_{T,i}$ such that:
\[
BS(w_i^TF_{t},K_i,T,V_{i,T})=p_i,\quad i=1,\ldots,m,
\]
where $p_i$ are the market prices of basket call options with weights
$w_i$, maturity $T$ and strike $K_i$. Note that we implicitly assume
that all the options have the same maturity, and that, without loss
of generality, the risk-free interest rate is zero (we compare prices
in the forward market). Let us remark however that the multiperiod
generalization has exactly the same format (see \cite{dasp02d}).

This last semidefinite program can be solved very efficiently
using algorithms such as the one by \cite{Stur99}, we refer the
reader to \cite{Nest94} or \cite{Boyd03} for further details. This
means that given a certain number of prices of liquid market
instruments, we can efficiently compute upper and lower bounds on
the model price of another instrument.

If the market price of an instrument falls outside of these bounds,
we then know that calibrating the model to this instrument together
with the original set of prices will be infeasible. This means that
there is a dynamic arbitrage opportunity if (and only if) the market
dynamics follow that of the model. In practice however, this more
often means either that the model dynamics are not rich enough to
capture the market price features or that there is a problem with the
market data set (liquidity, outliers, missing data, etc). In the next
section, we discuss ways of refining this diagnostic.

\section{Static arbitrage bounds}
In the previous section, we computed bounds on basket option
prices assuming that the underlying assets followed lognormal
dynamics. In this section, we are looking for price bounds on
options (given prices of other liquid options) without any
assumption on the asset distribution. The range of prices covered
will of course be much larger than in the lognormal case but
option prices falling outside of this range generate buy-and-hold
arbitrage opportunities, which are much more robust to liquidity
issues than the model arbitrage detailed in the last section.

Let $p \in \reals_+^m$, $K \in \reals_+^m$, $w \in \reals^n$, $w_i
\in \reals^n$, $i=1, \ldots, m$ and $K_0 \geq 0$. We consider here
the problem of computing upper and lower bounds on the price of an
European basket call option with strike $K_0$ and weight vector
$w_0$:
\BEQ
\label{e-static-bounds}
\BA{ll}
\mbox{min./max.} & \Expect_{\pi}(w_0^Tx-K_0)^+\\
\mbox{subject to}& \Expect_{\pi}(w_i^Tx-K_i)^+=p_i,\quad
i=1,\ldots,m,
\EA
\EEQ
with respect to all probability distributions $\pi$ with support
in $\reals^n_+$ on the asset price vector $x$, consistent with a
given set of observed prices $p_i $ of options on other baskets.
Note that we again implicitly assume that all the options have the
same maturity, and that, without loss of generality, the risk-free
interest rate is zero (we compare prices in the forward market).

We know from \cite{bert00} that the problem described in
(\ref{e-static-bounds}) is NP-Hard and in the next section we
describe several relaxation techniques providing upper (resp.
lower) bounds on the upper (resp. lower) bounds described by
(\ref{e-static-bounds}).

\subsection{Linear programming relaxation}
In dimension one we know that if $C(K)$ is a function giving the
price of an option of strike $K$, then $C(K)$ must be positive,
decreasing and convex. With $C(0)=S$, we have a set of
\textit{necessary} conditions on call prices for the absence of
arbitrage.

In fact it is well known (see \cite{Laur00} or \cite{bert00} among
others) that these conditions are also \textit{sufficient}, so
there is no arbitrage between some given market call prices $p_i$
if and only if there is a function C(K) such that:
\BIT
\item $C(K)$ positive
\item $C(K)$ decreasing
\item $C(K)$ convex
\item $C(K_i)=p_i$ and $C(0)=S,\quad i=1,\ldots,m$.
\EIT
These conditions are easily generalized to the multidimensional case
as follows: given a set of market prices for basket calls
$C(w_i,K_i)= p_i$ and suppose there is no arbitrage, then the
function $C(w,K)$ must satisfy:
\BIT
\item $C(w,K)$ positive
\item $C(w,K)$ decreasing in $K$, increasing in $w$
\item $C(w,K)$ jointly convex in $(w,K)$
\item $C(w_i,K_i)=p_i$ and $C(0)=S,\quad i=1,\ldots,m$.
\EIT
The key difference is here that these conditions are \emph{not
sufficient}. Nevertheless, we can form a relaxation of problem
(\ref{e-static-bounds})as:
\BEQ
\label{e-bounds-infinite}
\BA{ll}
\mbox{min./max.} & C(w_0,K_0)\\
\mbox{subject to}& C(w,K) \mbox{ positive}\\
& C(w,K) \mbox{ decreasing in $K$, increasing in $w$} \\
& C(w,K) \mbox{ jointly convex in $(w,K)$} \\
& C(w_i,K_i)=p_i \mbox{ and }C(w,0)=w^T S,\quad i=1,\ldots,m.
\EA
\EEQ
In this form, this is an infinite dimensional linear program and
not directly tractable. However, we can discretize
(\ref{e-bounds-infinite}) into a finite linear program by sampling
the constraints on the data points $(w_i,K_i)$. We get the
following program
\BEQ
\begin{array}{ll}
\mbox{maximize/minimize} & p_{0} \\
\mbox{subject to} & g_{i}^T((w_{j},K_{j})-(w_{i},K_{i})) \leq
p_{j}-p_{i},\quad i,j=0,\ldots,m \\
& g_{i,j}\geq 0,-1\leq g_{i,n+1}\leq 0,\quad i=0,...,m,\quad
j=1,\ldots,n\\
& g_{i}^T((w_{i},K_{i})) =p_{i},\quad i=0,\ldots,m,
\end{array}
\label{e-bounds-lp}
\EEQ
in the variables $p_0 \in \reals_+$ and $g_{i}\in \reals^{n+1}$
for $i=0,...,m$. This is a linear program with $m\times(n+1)$
variables and can be solved very efficiently. Furthermore,
\cite{dasp02c} show that program (\ref{e-bounds-lp}) is in fact an
\emph{exact discretization} of program (\ref{e-bounds-infinite})
and that an optimal solution to (\ref{e-bounds-infinite}) can be
constructed from that of (\ref{e-bounds-lp}).

This first relaxation technique allows us to get upper and lower
bounds on the solution to (\ref{e-static-bounds}) at a minimal
numerical cost, \cite{dasp02c} show that these bounds are exact in
some particular cases, but in general nothing can be guaranteed
about their performance.

Necessary and sufficient conditions describing when a function can be
written as the price
\[
p=\textstyle \Expect_{\pi}(w^Tx-K)^+
\]
of a basket call with weight $w$ and strike $K$ were derived by
\cite{henk90} in their investigation of production functions. They
get the following result: a function can be written
\[
C(w,K)=\int_{\mathbf{R}_{+}^{n}}(w^{T}x-K)^{+}d\pi (x),
\]
with $w\in \mathbf{R}_{+}^{n}$ and $K>0$, if and only if:
\BIT
\item  $C(w,K)$ is \textit{convex} and \textit{homogenous} of degree one
\item  for every $w\in \mathbf{R}_{++}^{n}$, we have
$\lim_{K\rightarrow \infty}C(w,K) = 0 \mbox{ and } \lim_{K\rightarrow
0^+} \frac{\partial C(w,K)}{\partial K} = -1$
\item   $F(w)=\int_0^{\infty }e^{-K}d\left( \frac{\partial C(w,K)}
{\partial K}\right)$ belongs to $C_{0}^{\infty }(\mathbf{R}_{+}^{n})$
\item  For some $\tilde{w}\in \mathbf{R}_{+}^{n}$,
$\left(-1\right)^{k+1}D_{\xi_1}...D_{\xi_k}F(\lambda\tilde{w})\geq
0,$ for all positive integers $k$ and $\lambda\in\mathbf{R}_{++}$ and
all $\xi _{1},\ldots,\xi _{k}$ in $\mathbf{R}_{+}^{n}$.
\EIT
The key difficulty here is that the last two smoothness and
monotonicity conditions are numerically intractable, so this
result is of very little practical help in refining the conditions
used in (\ref{e-bounds-lp}).

\subsection{A moment approach}
In this section, we look for ways of improving the relaxation
technique in (\ref{e-bounds-lp}). NP hardness means that we can't
hope to get a tractable set of necessary conditions to solve the
problem exactly, here instead we look for additional conditions on
prices that produce a sequence of successively tighter bounds on
the solution to (\ref{e-static-bounds}).

The integral transform approach above suggests a link to moment
theory. In fact, as detailed in \cite{dasp03c}, numerically
tractable conditions for the existence of a measure $\pi$ such
that $p=\textstyle \Expect_{\pi}(w^Tx-K)^+$ can be obtained by a
generalization of Bernstein-Bochner type results to the payoff
semigroup (see \cite{Berg84b} for a complete exposition). We
briefly recall this construction below.

We suppose that the market is composed of cash and $n$ underlying
assets $x_i$ for $i=1,\ldots,n$ with $x \in \reals_+^n$. We
suppose that the forward prices of the assets are known and given
by $p_i$, for $i=1,\ldots,n$, hence $w_i$ is the Euclidean basis
and $K_i=0$ for $i=1,\ldots,n$. In addition to these basic
products, there are $m+1$ basket \textit{straddles} on the assets
$x$, with payoff given by $|w_{n+i}^{T}x-K_{n+i}|$, $i=1, \ldots,
m$. Because a straddle is obtained as the sum of a call and a put,
we get the market price of straddles from those of basket calls
and forward contracts by call-put parity and the static arbitrage
problem on straddles is strictly equivalent to problem
(\ref{e-static-bounds}) since we always assume that forward prices
are quoted in the market.

For simplicity, we will note these payoff functions $e_i$, for
$i=0,\ldots,m+n$, with $e_{0}(x)=|w_0^{T}x-K_0|$, $e_i(x)=x_i$ for
$i=1,\ldots,n$ and $e_{(n+j)}(x)=|w_i^{T}x-K_i|$ for
$j=1,\ldots,m$. In what follows, we will focus on the commutative
semigroup $(\mathbb{S},\cdot)$ generated by the payoffs $e_i(x)$
for $i=0,\ldots,m+n$, the cash $1_{\mathbb{S}}$ and their
products.
\BEQ
\label{e-semigroup}
\mathbb{S}=\{1,x_1,\ldots,|w_m^{T}x-K_m|,x_1^2,\ldots,x_i|w_j^{T}x-K_j|,\ldots
\}
\EEQ

Let us recall that a function $f:\mathbb{S} \rightarrow \reals$ is
called positive semidefinite iff for all finite families $\{s_i\}$
of elements of $\mathbb{S}$, the matrix with coefficients $f(s_i
s_j)$ is positive semidefinite. We then get the following result
from \cite{dasp03c}, suppose the asset distribution has compact
support $K$ and $\mathbb{S}$ is the payoff semigroup defined
above. A function $f(s):\mathbb{S} \rightarrow \reals$ can be
represented as
\BEQ
f(s)=\textstyle \Expect_{\nu}[s(x)],\quad\mbox{for all
}s\in\mathbb{S}, \label{e-moment-constraints}
\EEQ
for some measure $\nu$ on $K$, and satisfies the price constraints
in (\ref{e-static-bounds}) if and only if:
\begin{enumerate}
\item[(i)] $f(s)$ is positive semidefinite,
\item[(ii)] $f(e_is)$ is positive semidefinite for $i=0,\ldots,n+m,$
\item[(iii)] $\left(\beta f(s) - \sum_{i=0}^{n+m}{f(e_is)}\right)$ is positive
semidefinite,
\item[(iv)] $f(e_{i})=p_i$ for $i=1,\ldots,n+m.$
\end{enumerate}
Furthermore, for each function $f$ satisfying conditions (i) to
(iv), the measure $\nu$. A similar result holds in the case where
the support of $\nu$ is not compact (see \cite{dasp03c}).

The above result shows that testing for the absence of static
arbitrage between the securities in $\mathbb{S}$, \ie~the set of
straddles and their products, is equivalent to testing the
positivity of an infinite number of matrices. This gives a direct
recipe for writing a relaxation to problem
(\ref{e-static-bounds}): by considering a finite subset of these
matrix inequalities, (\ref{e-static-bounds}) can be turned into a
semidefinite program. We summarize this procedure below.

We begin by recalling the construction of moment matrices in
\cite{Lass01}. Let $\mathcal{A}(\mathbb{S})$ denote the real
algebra generated by the payoffs in $\mathbb{S}$. We adopt the
following multiindex notation for monomials in
$\mathcal{A}(\mathbb{S})$:
\[
e^{\alpha}(x):=e_0^{\alpha_0}(x)e_1^{\alpha_1}(x)\cdots
e_{m+n}^{\alpha_{m+n}}(x),
\]
and we let
\BEQ
y_e=(1,e_0,\ldots,e_{m+n},e_0^2,e_0e_1,\ldots,
e_0^d,\ldots,e_{m+n}^d) \label{def:moment-vector-compact}
\EEQ
be the vector of all monomials in $\mathcal{A}(\mathbb{S})$, up to
degree $d$, listed in graded lexicographic order. We note $s(d)$
the size of the vector $y_e$. Let $y\in \reals^{s(2d)}$ be the
vector of moments (indexed as in $y_e$) of some probability
measure $\nu$ with support in $\reals^n_+$, we note $M_{d}(y)\in
\reals^{s(d)\times s(d)}$, the symmetric matrix:
\[
M_{d}(y)_{i,j}=\int_{\mathbf{R}^n_+}\left(y_{e}\right)_i (x)
\left(y_{e}\right)_j (x) d\nu(x),\quad \mbox{for }i,j=1,...,s(d).
\]
In the rest of the paper, we will always implicitly assume that
$y_1=1$. With $\beta (i)$ the exponent of the monomial
$\left(y_{\chi}\right)_i$ and conversely, $i(\beta)$ the index of
the monomial $e^{\beta}$ in $y_{e}$. We notice that for a given
moment vector $y\in \reals^{s(d)}$ ordered as in
(\ref{e-semigroup}), the first row and columns of the matrix
$M_{d}(y)$ are then equal to $y$. The rest of the matrix is
constructed according to:
\[
M_{d}(y)_{i,j}=y_{i(\alpha +\beta)}\mbox{ if
}M_{d}(y)_{i,1}=y_{i(\alpha)}\mbox{ and
}M_{d}(y)_{1,j}=y_{i(\beta)}.
\]

Similarly, let $g\in \mathcal{A}(\mathbb{S})$, we derive the
moment matrix for the measure $g(x)d\nu$ on $\reals_+^n$, noted
$M_{d}(gy)\in \symm^{s(d)}$, from the matrix of moments $M_{d}(y)$
by:
\[
M_{d}(gy)_{i,j}=\int_{\mathbf{R}^n_+}\left( y_{e}\right)_{i}(x)
\left(y_{e}\right)_{j}(x) g(x) d\nu(x)\quad \mbox{for
}i,j=1,...,s(d).
\]
The coefficients of the matrix $M_{m}(gy)$ are then given by:
\BEQ
M_{d}(gy)_{i,j}=\sum_{\alpha}g_{\alpha}y_{i(\beta
(i)+\beta(j)+\alpha)} \label{def:localizing-matrix-compact}
\EEQ

We can now form a semidefinite program to compute a lower bound on
the optimal solution to (\ref{e-static-bounds}) using a subset of
the moment constraints in (\ref{e-moment-constraints}), taking
only monomials and moments in $y$ up to a certain degree. Let $N$
be a positive integer and $y\in\reals^{s(2N)}$, a lower bound on
the optimal value of:
\[
\BA{ll}
\mbox{minimize} & {p_0:=\Expect}_{\nu }[e_0(x)] \\
\mbox{subject to} & {\Expect}_{\nu }[e_i(x)]=p_i, \quad
i=1,\ldots,n+m,
\EA
\]
can be computed as the solution of the following semidefinite
program (again see \cite{Nest94} or \cite{Boyd03}):
\BEQ
\BA{ll}
\mbox{minimize} & y_2 \\
\mbox{subject to} & M_N(y) \succeq 0 \\
 & M_N(e_jy) \succeq 0 , \quad \mbox{for }j=1,\ldots,n,\\
 & M_N\left((\beta-\sum_{k=0}^{n+m}{e_k})y)\right) \succeq 0 \\
 & y_{(j+2)}=p_j, \quad \mbox{for }j=1,\ldots,n+m \mbox{ and }
 s\in \mathbb{S}\\
\EA
\label{eq:sdp-bound-compact}
\EEQ
where $s$ is such that $i(s)\leq s(2N)$. As $N$ increases, program
(\ref{eq:sdp-bound-compact}) provides an increasingly precise
relaxation of problem (\ref{e-static-bounds}).

\section{Numerical results}
Here, we try to quantify on simple examples the magnitude of the
gap between model prices, the price range obtained using the
various relaxation techniques detailed above and the exact price
bounds in (\ref{e-static-bounds}).

\subsection{Discrete model} Here, we simulate a set of
arbitrage free basket call prices using a simple discrete model.
Given these prices and the absence of arbitrage between basket
calls, we study the price bounds induced on another basket call.

Of course we can't compare these bounds with the exact solution,
however we can try to compare these bounds with \textit{inner}
bounds obtained by maximizing and minimizing the price
$C(\omega_0,K_0)$ over a set of probability measures satisfying
the price constraints $C(\omega_i, K_i)=p_i$. If we consider only
discrete measures, this becomes a (very large) linear program.

We suppose here that the asset price at maturity $T$ lies within
the unit box $[0,1]^n$. We then discretize the probability density
using a grid with $N$ bins per asset. The problem of finding
(inner) upper and lower bounds on a basket $(\omega_0,K_0)$ can be
written as:
\BEQ
\label{e-cons-inner-example}
\BA{ll}
\mbox{max./min.}   & \Expect_\nu \left(\omega_0^Tx-K_0\right)_+ \\
\mbox{subject to}  & \Expect_\nu \left(\omega_i^Tx-K_i\right)_+=p_i,\quad i=1,\ldots,m, \\
\EA
\EEQ
which is a linear program of (exponential) size $N^n$ in the
discrete measure $\nu$.

\begin{figure}[p]
\begin{center}
\psfrag{str}[t][b]{strike}
\psfrag{price}[b][t]{price}
\psfrag{bnds}[c][t]{Price bounds}
\includegraphics[width=0.62 \textwidth]{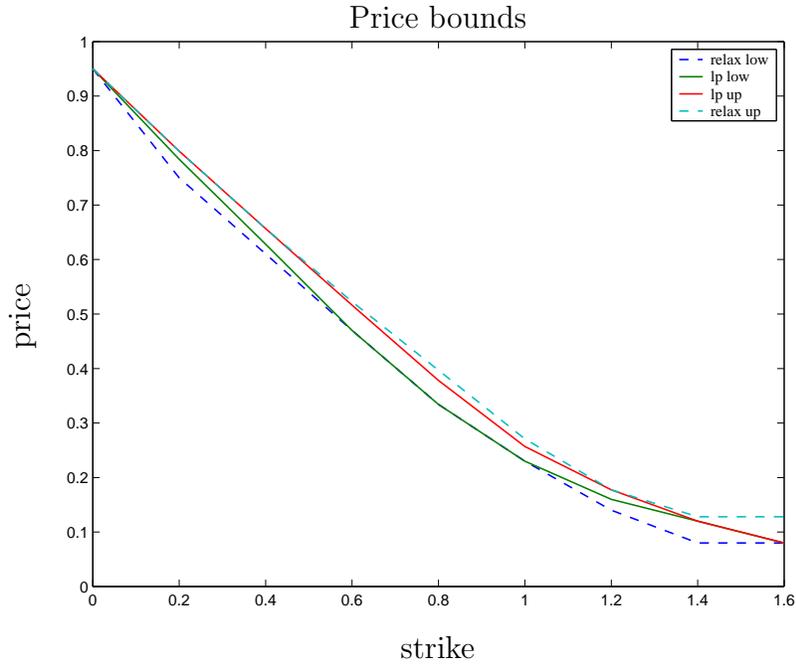}
\caption{Comparison of the inner bounds computed by discretization
(solid lines, computed using~(\ref{e-cons-inner-example})) and the
outer bounds obtained by relaxation (dashed lines, computed
using~(\ref{e-bounds-lp})).\label{fig-inner-outer}}
\end{center}
\end{figure}

We test these upper and lower bounds in dimension two.
Program~(\ref{e-cons-inner-example}) can be written:
\[
\BA{ll}
\mbox{max./min.}   & \sum_{k,l=0,\ldots,N} \nu_{kl}\left(\omega_0^T[k/N,l/N]^T-K_0\right)_+ \\
\mbox{subject to}  & \sum_{k,l=0,\ldots,N} \nu_{kl}\left(\omega_i^T[k/N,l/N]^T-K_i\right)_+=p_i,\quad i=1,\ldots,m.\\
\EA
\]
which is a linear program in the variable $\nu\in\reals^{N\times
N}$. The assets are noted $x_1,x_2$ and we look for bounds on the
price of an index option with payoff $(x_1+x_2-K)_+$. To produce
price data, we use a simple discrete model for the assets, their
distribution has finite support and is given by:
\[
x =\{ (0,0) ,(0,.8) , (.8,.3) ,(.6,.6) , (.1,.4) , (1,1)\}
\]
with probability
\[
 \{.2 , .2 , .2 , .1 , .1 ,.2\}.
\]
The input data set is composed of the forward prices together with
the following call prices:
\[
\BA{l}
(.2x_1+x_2-.1)_+,~(.5x_1+.8x_2-.8)_+,~(.5x_1+.3x_2-.4)_+,\\
(x_1+.3x_2-.5)_+,~(x_1+.5x_2-.5)_+,~(x_1+.4x_2-1)_+,~(x_1+.6x_2-1.2)_+.
\EA
\]
We plot the inner and outer bounds obtained using this data in
figure~(\ref{fig-inner-outer}). We observe that sometimes the
bounds match, \ie~the price bounds given by the relaxation are
tight, while sometime there is a gap and not much can be said
about the relaxation's suboptimality.

\begin{figure}[p]
\begin{center}
\psfrag{instruments}[t][b]{number of instruments}
\psfrag{price}[b][t]{price}
\psfrag{bnds}[c][t]{Price bounds}
\includegraphics[width=0.6 \textwidth]{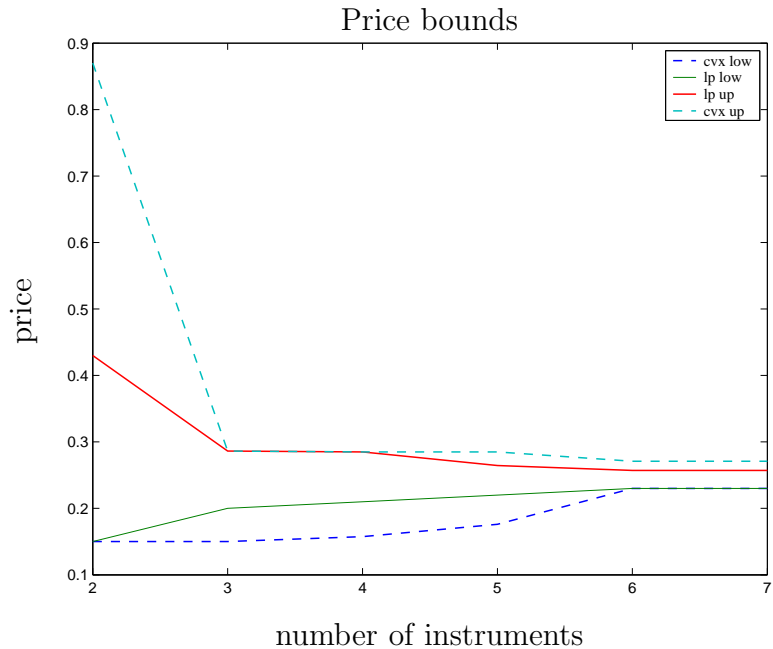}
\caption{Inner bounds (solid lines, computed
using~(\ref{e-cons-inner-example})) and outer bounds (dotted
lines, computed using~(\ref{e-bounds-lp})) versus number of
instruments in the data set.\label{fig-inner-outer-conv}}
\end{center}
\end{figure}

We also examine how these bounds evolve as more and more
instruments are incorporated into the data set. For a particular
choice of strike price (here $K=1$), we compute the outer
bounds~(\ref{e-bounds-lp}) and inner
bounds~(\ref{e-cons-inner-example}) obtained when using only the
$k$ first instruments in the data set, for $k=2,\ldots,7$. The
result is plotted in figure~(\ref{fig-inner-outer-conv}). We
notice that it takes relatively few basket prices to get good
bounds on the price of the target option.

\subsection{Multivariate lognormal model} Here again, we simulate a set of
arbitrage free basket call prices using this time the multivariate
\cite{Blac73} model detailed in section one. Given these prices
and the absence of arbitrage between basket calls, we study the
price bounds induced on another basket call.

We look for bounds on the price of an index option:
$(.2\sum_{i=1}^5x_i-K)_+$, given the price of all at-the-money
single asset calls and at-the-money basket calls with the
following weights:
\[\left[
\BA{ccccc}
0.33 & 0.33 & 0.33 & 0.00 & 0.00 \\
0.00 & 0.00 & 0.33 & 0.33 & 0.33 \\
0.40 & 0.20 & 0.20 & 0.20 & 0.00 \\
\EA
\right]\]

The assets initial values $F_t^i$ are $(0.03,0.03,.05,.07,.07)$
and the model covariance matrix $X$ is given by:
\[\left(
\BA{ccccc}
0.06 & 0.04 & 0.04 & 0.04 & 0.04 \\
0.04 & 0.06 & 0.04 & 0.04 & 0.04 \\
0.04 & 0.04 & 0.06 & 0.04 & 0.04 \\
0.04 & 0.04 & 0.04 & 0.06 & 0.04 \\
0.04 & 0.04 & 0.04 & 0.04 & 0.06 \\
\EA
\right)\]

\begin{figure}[p]
\begin{center}
\psfrag{str}[t][b]{strike}
\psfrag{price}[b][t]{price}
\psfrag{bnds}[c][t]{Price bounds}
\includegraphics[width=0.60 \textwidth]{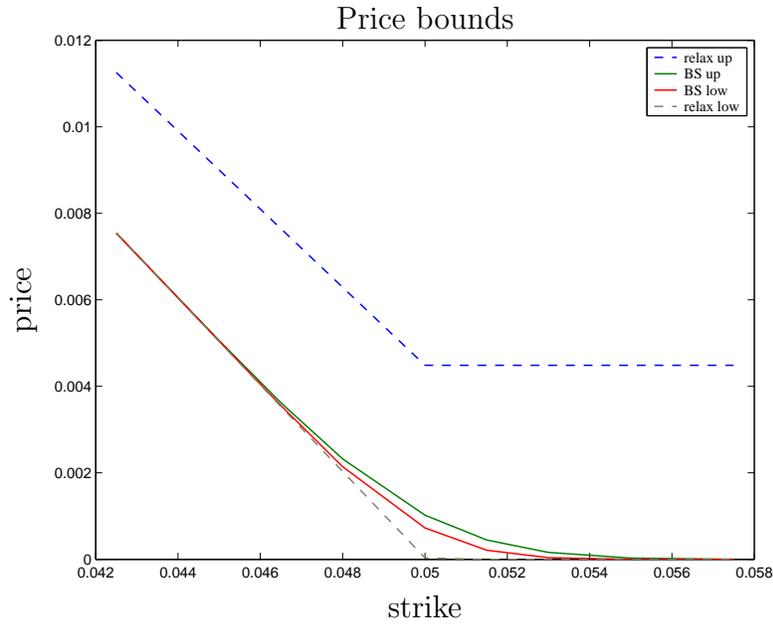}
\caption{Comparison of the inner bounds computed using the Black
\& Scholes model (solid lines, computed using~(\ref{e-sdp-bs}))
and the outer bounds obtained by relaxation (dashed lines,
computed using~(\ref{e-bounds-lp})).\label{fig-inner-outer-bs}}
\end{center}
\end{figure}

We plot the inner and outer bounds obtained using this data in
figure~(\ref{fig-inner-outer-bs}). Since the \cite{Blac73} of an
at-the-money call is not very sensitive to the volatility, we
could have expected the call price near the money to be somewhat
insensitive to model specification. The fact that the gap between
the inner model price bounds and the outer static arbitrage bounds
is so large seems to directly contradict this intuition, showing
that in fact the range of arbitrage free prices for basket calls
is much larger than the range of prices that can be attained by a
multivariate \cite{Blac73} model.

\subsection{Moment constraints}
\begin{figure}[p]
\begin{center}
\psfrag{strk}[t][b]{strike}
\psfrag{pr}[b][t]{price}
\psfrag{tit}[c][t]{Price bounds on a straddle}
\includegraphics[width=0.6 \textwidth]{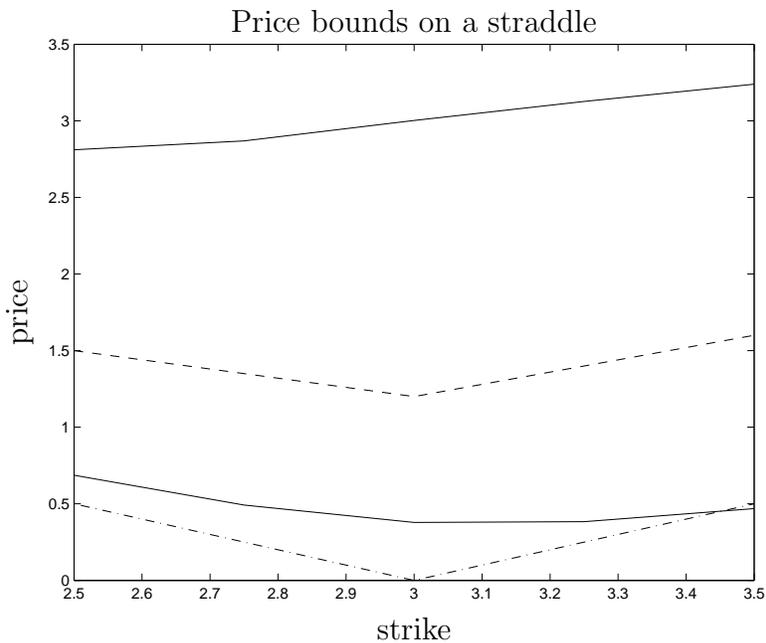}
\end{center}
\caption{Upper and lower price bounds on a straddle (solid lines).
The dashed lines represent the payoff function and the actual
price. \label{fig-harmonic-bounds}}
\end{figure}

For numerical reasons, we consider a model with two assets
$x_1,x_2$ and look for bounds on the price of the basket
$|x_1+x_2-K|$. We use a simple discrete model for the assets:
\[x=\{(0,0),(0,3),(3,0),(1,2),(5,4)\}\],
with probability
\[p=(.2,.2,.2,.3,.1)\],
to simulate market prices for the forwards and the following
straddles:
\[
|x_1-.9|,|x_1-1|,|x_2-1.9|,|x_2-2|,|x_2-2.1|.
\]
The results are detailed in figure~(\ref{fig-harmonic-bounds}).
Unfortunately the code from \cite{Stur99}, although very robust in
general, is not as stable on the moment problems detailed here. A
different implementation using large scale optimization techniques
such as spectral bundle should prove more appropriate here.

\section{Conclusion}
We have detailed some tractable relaxation procedures to test for
the absence of static arbitrage between basket calls. A comparison
with the range of prices attained by a multivariate \cite{Blac73}
model shows that this simple model only covers a relatively small
portion of the range of arbitrage free prices. Running the same
kind of test on a model with a richer smile structure, if
numerically feasible, would be very interesting. Also, the
numerical issues encountered in the moment relaxation technique
are very surprising and probably deserve more investigation.

\bibliographystyle{amsalpha}
\bibliography{MainPerso}

\end{document}